# Anti-Thermal Quenching Phosphors based on Metal Halides

Baowei Zhang,[a,*] Liberato Manna[b,*]

[a] College of Chemistry, Zhengzhou University, Zhengzhou, Kexue road, 450001 China

[b] Nanochemistry, Istituto Italiano di Tecnologia, Via Morego 30, 16163, Genova, Italy

**ABSTRACT:** Thermal quenching (TQ) generally occurs in phosphors and is ascribed to the activation of non-radiative transitions at elevated temperatures. This effect limits the use of most phosphors in high-power/high-temperature applications, such as outdoor lighting and laser systems. To achieve anti-TQ properties, structural design of phosphors is required. This usually follows two guidelines: (1) increasing lattice rigidity to minimize thermal expansion; (2) converting thermal energy into radiative transitions to compensate for the non-radiative losses. While metal oxides and metal nitrides dominate the field of commercial anti-TQ phosphors, metal halides—despite their inherently soft lattices—have shown remarkable progress as anti-TQ phosphors in recent years. Here, we review the advances in anti-TQ metal halides (covering the time span from 2017, when first reports have appeared, till today) and discuss their mechanisms and applications. We argue that the low synthesis temperatures of metal halides and their high photoluminescence quantum yields (PLQYs) make them promising candidates as anti-TQ phosphors. Furthermore, since the rich optical-physical processes underlying the anti-TQ effect in soft-lattice in metal halides are only now beginning to be unraveled, this creates opportunities for many fundamental investigations.

High luminous power and brightness is essential for outdoor lighting that significantly extends human productive/social time[1]. Yet, in most phosphors high-power operation raises device temperatures up to 200°C[2], triggering thermal quenching (TQ) of the emission and leading to energy waste. To limit the losses in the performance and efficiency of high-power lighting devices, two strategies are employed: (1) external thermal management by enhancing heat dissipation with the use of heat sinks or high-thermal-conductivity media, to limit increases in device temperature under high power operation conditions[3]; (2) intrinsic phosphor design, that is, deployment of anti-TQ phosphors as emitters[4]. This second strategy will be the focus of this review, which will highlight recent developments on anti-TQ phosphors based on metal halides.

Fundamentals studies on anti-TQ phosphors date back to the 1960s, when G. Blasse established the theoretical framework for metal oxide phosphors doped with lanthanide (Ln) ions[5-7]. In the top three diagrams of Figure 1, the blue upward arrow represents the absorption process, while the green downward arrow denotes the emission process. TQ occurs when thermal excitation (red zigzag line) promotes the system from the excited-state minimum (B) to the ground/excited-state crossover point (C). As sketched in Figure 1a and 1d, complete TQ occurs at any given temperature when C is the global minimum of excited states. As shown in Figure 1b and 1e, TQ is possible even without the assistance of thermal energy (E=kT, where k is the Bolzman constant and T is the environment temperature), when C lies above the minimum point of the excited state curve (point B), but below the energy level reached after absorption (point A). As shown in Figure 1c and 1f, anti-TQ occurs when C lies above A, that is, when the thermal energy is not enough to surmount the energy barrier (ΔE).

Mainstream anti-TQ systems are based on metal oxides and metal nitrides[8,9]. The underlying feature of these materials is their rigid structure characterized by a small thermal-lattice expansion coefficient[2,5,10]. For example, Kim et al. reported in 2017 a blue-emitting $Na_{3-2x}Sc_2(PO_4)_3$:xEu$^{2+}$ phosphor with zero-thermal quenching emission at 200 °C[2], and Qiao et al. developed in 2018 a $K_2BaCa(PO_4)_2$:Eu$^{2+}$ phosphor that preserved 100% of its room temperature PL intensity at 275 °C[11]. Metal halides, on the other hand, represent a broad class of materials characterized by $[M^{m+}X_n]^{m-n}$ polyhedral skeletons (M = metal, X = Cl, Br, I) that are often charge-stabilized by A$^+$ cations (A$^+$=K$^+$, Rb$^+$, Cs$^+$, or an organic cation)[12]. Metal halides are generally not considered as anti-TQ phosphors, as they tend to have "soft lattices"[13]. Nevertheless, they have gained increasing attention in recent years (Table 1), following the discovery by Yuan et al. in 2017 of anti-TQ behavior from Mn$^{2+}$ doped CsPbCl$_3$ relatively to the emission from Mn$^{2+}$ states[14]. Anti-TQ metal halides fall into three categories: (1) metal halides doped with metal ions except Ln ones (for example Mn$^{2+}$, Sb$^{3+}$, Mo$^{4+}$), in which the thermally assisted energy transfer from host/defects to the emission center can compensate the energy losses of the emission center, so that the phosphor can reach stable emission even at high temperature[4]; (2) metal halides doped with Ln ions, in which the 4f-4f transitions of the Ln elements are insensitive to the coordination environment, so they preserve their emission characteristics under thermal lattice expansion[15,16]; (3) metal halides that have undergone a surface treatment (so far documented only for lead halide perovskite nanocrystals) to form a wide bandgap surface layer, allegedly leading to a type-I core-shell structure that can mitigate the TQ process[17, 18]. Research on anti-TQ phosphors based on metal halides is a relatively new field of research, born only eight years ago and now experiencing a boost. This field is not yet covered by any review so far, hence the rationale for this focus review[19].

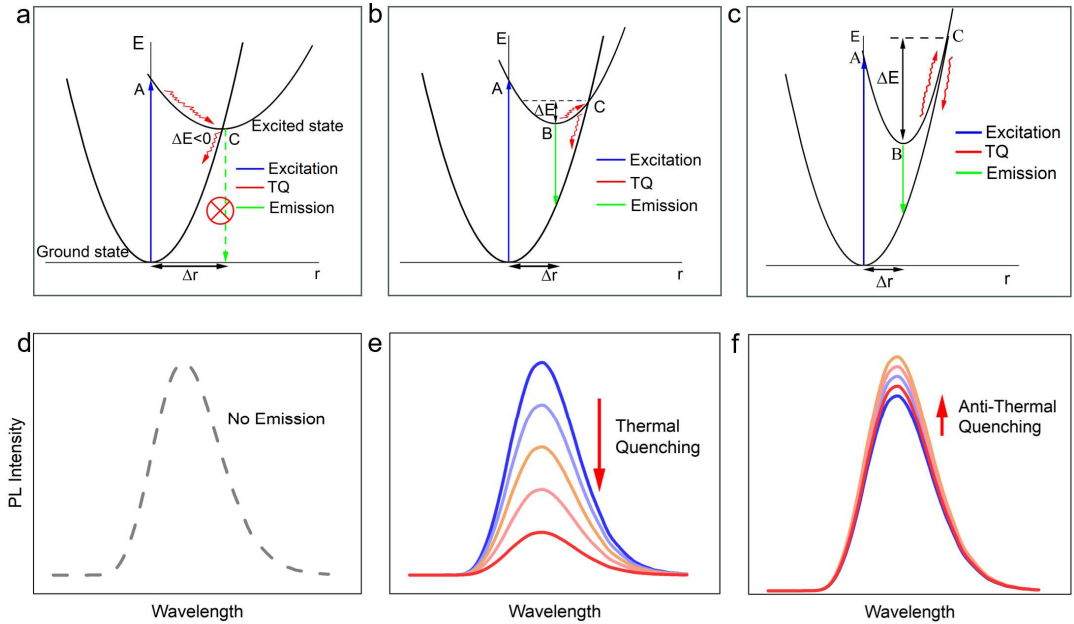

**Figure 1.** Three different configuration coordinate diagrams (a-c) for thermal quenching processes and their corresponding thermal quenching behavior (d-f), based on the model proposed by G. Blasse in 1974[7]. Reproduced (Adapted) with permission from [7]. Copyright 1974, Elsevier. In Figure a-c, r is the configuration coordinate representing the average position of the atoms surrounding a luminescent center, while Δr represents the difference between the equilibrium distance of the ground and excited state.

| Material class | Composition | Synthesis temperature (°C) | Emission peak (nm) | PLQY (%) | $T_c$ (K) |
|---|---|---|---|---|---|
| Metal halides | [20]$CsCdCl_3$:$x$%$Br^-$ | 180 | 482 | 84 | 377 |
| | [4]$Rb_3InCl_6$:$x$%$Sb^{3+}$ | 110 | 521 | 90 | 500 |
| | [21]$Rb_3(Cd_{0.8}Mn_{0.2})_2Cl_7$ | 150 | 575 | 88 | 423 |
| | [22]$Cs_2ZnCl_4$:$x$%$Sb^{3+}$ | 150 | 745 | 75 | 500 |
| | [23]$Cs_2MoCl_6$ | 180 | 1000 | 26 | 473 |
| | [24]$CsPbCl_3$: $x$%$Yb^{3+}$/$Er^{3+}$ | 240 | 1540 | 4 | 360 |
| Metal oxides/nitrides /fluorides | [2]$Na_{3-2x}Sc_2(PO_4)_3$:$x$$Eu^{2+}$ | 1300 | 453 | 74 | 473 |
| | [25]$MgAl_2O_4$: $Mn^{2+}$/$Eu^{2+}$ | 1200 | 516 | - | 423 |
| | [26]$K_2SiF_6$:$Mn^{4+}$ | 25* | 630 | 90 | 448 |
| | [27]$Lu_2SrAl_4SiO_{12}$:$Cr^{3+}$ | 1500 | 710 | 77 | 573 |

**Table 1.** Composition, synthesis temperature (°C), emission peak (nm), PLQY (%) and $T_c$ (K) for metal halides and metal oxides/nitrides anti-TQ phosphors covering the emission spectrum from the whole visible range to the near infrared (NIR). $T_c$ is defined as the temperature at which the PL intensity is equal to 100% of the initial PL intensity at 300 K. $T_c$ can be used as one of the parameters characterizing the anti-TQ properties of a material.

### 1. Mechanism Discussion for Anti-TQ metal halides

The design principles of anti-TQ phosphors follow two main avenues: (1) Minimizing the emission losses caused by thermally induced transitions to the crossover point of ground/excited state; (2) Introducing external energy to compensate the emission losses of the emission center. based on these principles, four alternative mechanisms have been proposed to explain the anti-TQ behavior in metal halides. The first one is the creation of an extra energy level in the band structure (Figure 2a). This energy level can be in the form of a defect (either from the host or from another dopant). At high temperature, energy transfer from this extra energy level to the emission center is activated, providing additional carriers to the emission center, compensating the non-radiative losses[2, 11]. This compensation mechanism is adopted by most Mn doped[21], hetero-valent atom doped[28] and multielement-doped[16] metal halides.

Also, if the lattice exhibits high structural rigidity, that it, if the lattice expansion is minimal or even negative at high temperatures, both structure and emission intensity of the emission center will be preserved and the material will exhibit anti-TQ behavior. This high lattice rigidity is common among anti-TQ metal oxide/nitride phosphors[29, 30]. For example, some of $A_2M_3O_{12}$ type anti-TQ phosphors exhibit zero-thermal expansion or even negative thermal expansion up to 500 K[8, 31]. However, most metal halides do not have comparable structural rigidity (they have a so-called soft lattice), due to their low lattice formation energy[32]. Instead, zero-dimensional (0D) metal halides, with their structure characterized by isolated polyhedra, appear to have a local structural rigidity: at high temperature, although these 0D metal halides undergo significant thermal expansion globally, this is mainly accounted for by the elongation of the inter-polyhedral distances, while the individual polyhedra (which are also the emission centers) are only marginally expanded (Figure 2b)[4]. Thanks to this local structural rigidity, 0D metal halides have been found to exhibit better anti-TQ performance than their 3D counterparts, as reported in Table 2[4, 33].

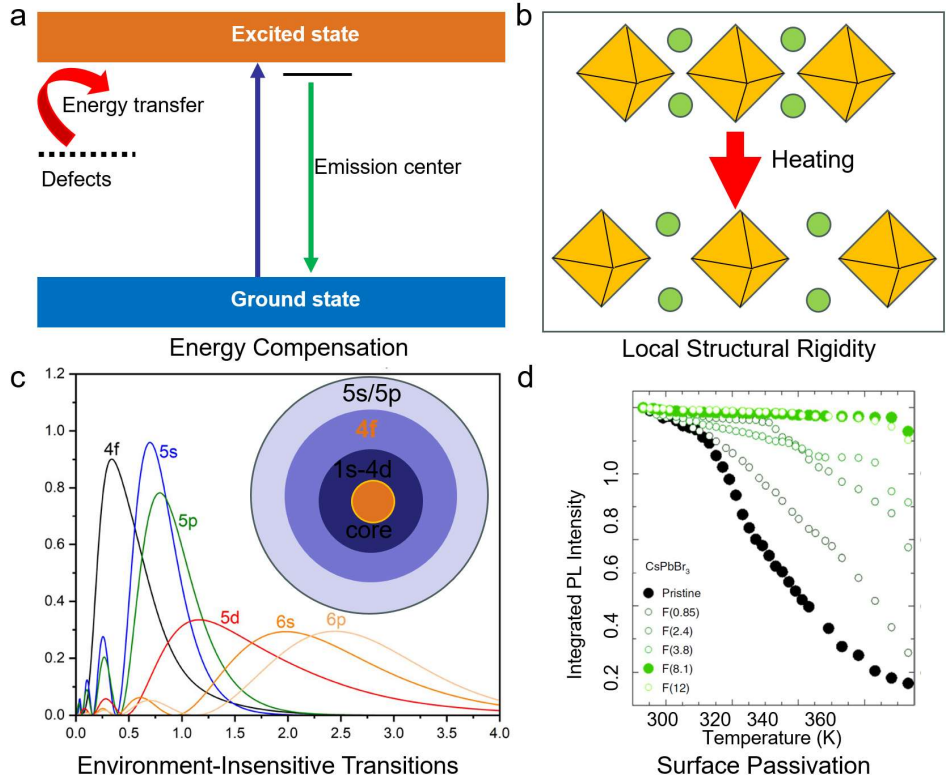

**Figure 2:** A summary of anti-thermal (anti-TQ) quenching mechanisms in some emissive metal halides. (a) anti-TQ arising from thermally activated energy transfer from defect (either from the host or from another dopant) to emission center, compensating the emission loss at high temperature; (b) anti-TQ due to local structural rigidity of 0D metal halides, preventing major changes in the emission center during the lattice expansion at high temperature; (c) anti-TQ arising from environment-insensitive 4f-4f transitions, as f orbitals are very contracted (the radial distribution function of 4f states is peaked much closer to the nucleus compared to 5s, 5p and 6s states), therefore they are not engaged in bonding with neighboring orbitals[37, 38]. Reproduced (Adapted) with permission from [37]. Copyright 2023, American Chemical Society; (d) anti-TQ arising from surface passivation of lead halides perovskite nanocrystals that suppresses the thermally activated non-radiative processes[17]. Reproduced (Adapted) with permission from [17]. Copyright 2021, Springer Nature.

| Materials class | Compositions | Dimensionality | Emission peak (nm) | PLQY (%) | $T_c$ (K) |
|---|---|---|---|---|---|
| 0D and 3D Sb doped[4] | $Rb_3InCl_6$:x%$Sb^{3+}$ | 0D | 521 | 90 | 500 |
|  | $Cs_2AgInCl_6$:x%$Sb^{3+}$ | 3D | 650 | 51 | <300 |
| 0D, 1D and 3D Ln halide[33] | $C_{72}N_{12}[Eu_2I_9]_2Eu_4I_{15}$ | 0D | 450 | 95 | 433 |
|  | $C_{16}N_4Eu_8I_{24}$ | 1D | 450 | 90 | 353 |
|  | $CsEuCl_3$ | 3D | 405 | 85 | 313 |

**Table 2**: Composition, dimensionality, emission peak (nm), PLQY (%) and $T_c$ (K) for 0D metal halides and their 3D counterparts[4, 33].

In the cases of Ln elements, their 4f-4f transitions are essentially environment insensitive, as f orbitals, being very contracted, do not engage in bonding with the orbitals of neighboring atoms (Figure 2c). Most Ln ion-doped metal halides show some degree of anti-TQ[16, 33]. Surface treatment is also an emerging strategy to improve the optical properties of lead halide perovskite nanocrystals[17, 34-36], and recent findings have revealed how a good surface passivation can efficiently suppress non-radiative processes even at high temperatures. For example, by post-treatment with dodecyl dimethylammonium fluoride (DDAF), $CsPbBr_3$ nanocrystals at 373 K preserves 90% of their room temperature PL (Figure 2d)[17].

In the following sections, we will divide the metal halides into three types: (1) metal halides doped with various metal ions (for example $Mn^{2+}$, $Sb^{3+}$, $Mo^{4+}$). We start with this class of materials as historically they were the first ones to be investigated; (2) metal halides doped with Ln ions; (3) surface treated lead halides nanocrystals, and discuss their anti-TQ behavior case by case.

## 2. Anti-TQ Metal Halide with Metal Ions Doping

**2.1. $Mn^{2+}$ doped cases.** $Mn^{2+}$ doping was the firstly reported strategy to achieve anti-TQ metal halide phosphors. As anticipated in the introduction, in 2017 Yuan et al. synthesized $Mn^{2+}$ doped $CsPbCl_3$ nanocrystals and investigated their temperature dependent PL behavior. The PL intensity of $Mn^{2+}$ increased with heating, an effect that was attributed to the accelerated energy transfer from the host to $Mn^{2+}$ dopant states, indicating a mild anti-TQ behavior from 60 K to 300 K[14]. In another work, Pinchetti et al. observed that, in $Mn^{2+}$ doped $CsPbCl_3$ nanocrystals, the $Mn^{2+}$ emission exhibits anti-TQ behavior in the 60-300 K range and TQ behavior in the 3.5-60 K range, see Figure 3a[39]. This anti-TQ behavior was explained by a two-step sensitization pathway: energy transfer from excitonic states to trap states and then to $Mn^{2+}$ dopant states (Figure 3b). These initial temperature dependent PL studies in the low temperature range uncovered the potential for $Mn^{2+}$ doped $CsPbX_3$ as anti-TQ phosphors.

In 2018, Ji et al. investigated the temperature dependent PL of $Mn^{2+}$ doped $CsPbCl_3$ nanocrystals with sizes from 5.0 to 17 nm in a 280-400 K temperature range[40]. They demonstrated that the anti-TQ behavior is size dependent. The maximum PL intensities from $Mn^{2+}$ states were observed at 260, 260, 280, 300, and 340 K for nanocrystals with sizes of 5.0, 7.2, 8.7, 13 and 17 nm, respectively. The authors claimed that the reduced surface trap density in larger nanocrystals might contribute to the suppression of non-radiative processes at high temperatures. In 2022, Kim et al. reported the synthesis of $Rb_3(Cd_{1-x}Mn_x)_2Cl_7$ microcrystals exhibiting zero-TQ up to 150 °C[21]. As shown in Figure 3c, the PL intensity was not quenched and was only slightly blue shifted at increasing temperature, from 25 °C to 150 °C. This anti-TQ property was attributed to thermally assisted energy transfer from defect states to luminescent centers. The activation energy of such defects states was estimated to be around 0.7-0.8 eV by thermoluminescence (TL) measurements, based on the formula $E_a$ (eV) = $T$(K)/500, where $E_a$ represents the activation energy and $T$ is the peak position in the TL spectra[41]. In 2023, Liu et al. prepared $Mn^{2+}$ doped $Cs_2CdCl_4$ crystals with Ruddlesden-Popper phase[42]. This phosphor had a long-persistent luminescence with decay lifetimes ranging from 450 s to 600 s, and exhibited anti-TQ PL in a 300-500 K temperature range. In the same year, Zhou et al. prepared $Mn^{2+}/Zr^{4+}$ co-doped $CsCdCl_3$ crystals featuring zero-TQ radioluminescence up to 448 K under X-ray excitation. In that system, doping with hetero-valent ions ($Zr^{4+}$) introduces structural defects in the lattice, which can then transfer energy to the $Mn^{2+}$ dopants at elevated temperature (Figure 3d)[43]. In 2024, Wu et al. developed a hybrid organic-inorganic $(Gua)_2MnBr_4$ (Gua = N, N'-diphenylguanidinium) single crystals that also exhibited anti-TQ behavior up to 400 K under X-ray excitation[44]. Such anti-TQ behavior was found to be excitation dependent: the anti-TQ was observed when the samples were excited by 405 nm light (indirect excitation), while TQ was observed when excited by 365 nm light (direct excitation), as shown in Figure 3e. Through molecular dynamic simulations, the authors claimed that the Mn-ligands distance increases with increasing temperature, while the Mn-Mn distance in the lattice surprisingly decreases at increasing temperature. This abnormal shortening of the Mn-Mn distance was believed to contribute to the anti-TQ photoluminescence behavior of the $(Gua)_2MnBr_4$ crystals.

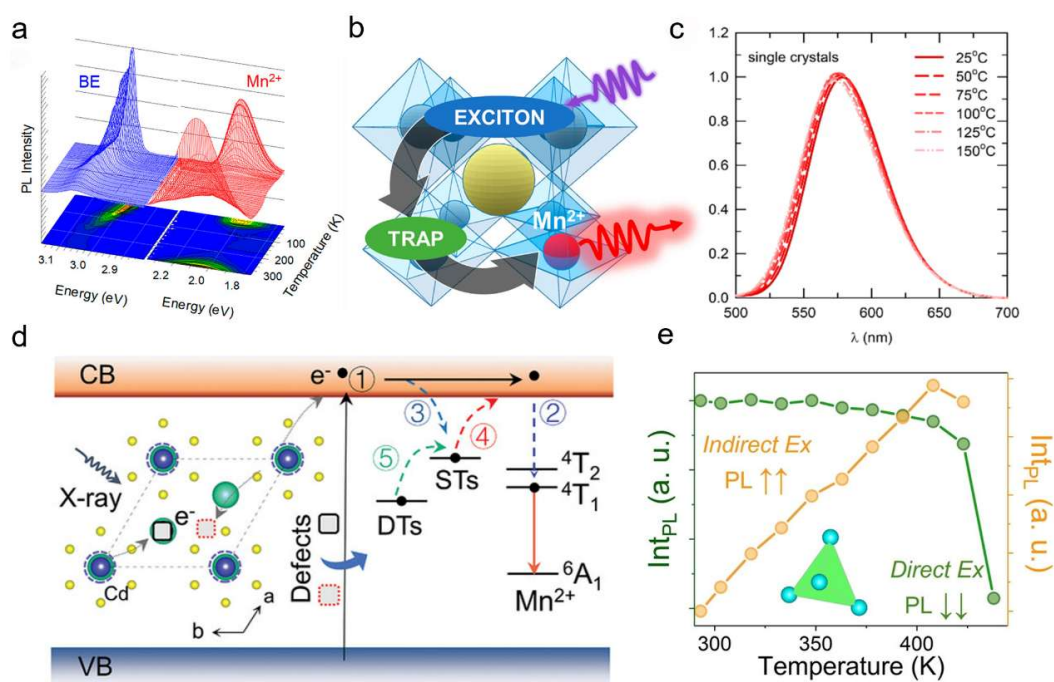

**Figure 3. $Mn^{2+}$ based anti-TQ metal halides.** (a) Temperature dependent dual-emission of $Mn^{2+}$ doped $CsPbCl_3$ nanocrystals at 3.5-300 K and (b) the proposed two-step sensitization model[39]. Reproduced (Adapted) with permission from [39]. Copyright 2019, American Chemical Society; (c) Temperature dependent PL spectra of $Rb_3(Cd_{1-x}Mn_x)_2Cl_7$ microcrystals at 25-150 °C[21]. Reproduced (Adapted) with permission from [21]. Copyright 2022, American Chemical Society; (d) A simplified model to describe the zero-TQ processes of the $CsCdCl_3$: 5%$Mn^{2+}$, 0.1%$Zr^{4+}$ powders, including the formation of excitons (①), radioluminescence (①②), trapping of electrons (③), de-trapping of electrons/thermal ionization (④) and energy transfer (⑤)[43]. Reproduced (Adapted) with permission from [43]. Copyright 2023, John Wiley and Sons; (e) Excitation-dependent anti-TQ behavior of $(Gua)_2MnBr_4$ single crystals[44]. Under 360 nm excitation, the phosphor shows TQ behavior. Conversely, when excited at 405 nm, it shows anti-TQ behavior. Reproduced (Adapted) with permission from [44]. Copyright 2025, John Wiley and Sons.

**2.2. $Sb^{3+}$ doped cases.** $Sb^{3+}$ is another widely used dopant in anti-TQ metal halide phosphors. $Mn^{2+}$ halides usually exhibit fixed orange-red emission (when six-coordinated) or green emission (when four-coordinated), while emission from $Sb^{3+}$ halides can be tuned from the visible range to the NIR by adjusting its coordination environment[45]. In 2022, Li et al. reported the synthesis of $Sb^{3+}$-doped $(BTPP)_2MnCl_4$ (BTPP = Benzyl triphenyl phosphonium) single crystals exhibiting dual emission: self-trapped exciton (STE) emission from $[SbCl_4]^-$ and an emission due to the $^4T_1$-$^6A_1$ transition from $Mn^{2+}$ states. The STE emission preserved 72.5% of its 298 K PL intensity at 420 K (Figure 4a). Such TQ resistance was attributed to the efficient energy transfer from the $[MnCl_4]^{2-}$ host to $[SbCl_4]^-$, promoting STE emission at elevated temperatures[46]. In 2024, Zhang et al. investigated the temperature dependent PL spectra of $Sb^{3+}$ doped $Cs_2ZnCl_4$ crystals in the 80–480 K range. Green emission (550-565 nm) was ascribed to the $[ZnCl_4]^{2-}$ host (STE-2), while NIR emission (690-780 nm) originated from the $[SbCl_4]^-$ dopant (STE-3). As the temperature increased, the STE-2 emission gradually vanished while the STE-3 emission intensified between 80-220 K and slightly decreased between 220-380 K, indicating a mild anti-TQ behavior (Figure 4b). The authors proposed that the lattice softening at high temperature promotes the formation of STEs, thereby enhancing the luminescence intensity[28].

In 2024, Zhang et al. (our groups) reported $Rb_3InCl_6$:$xSb^{3+}$, in the form of powders, as a robust anti-TQ phosphor exhibiting zero-TQ emission up to 500 K[4]. This property appears to stem from the combination of a defect compensation effect and an intrinsic structural rigidity of the isolated octahedra in the 0D structural framework of this material. First, the creation of intentional $InCl_3$ deficiencies during the synthesis generated a high density of lattice defects. Defect energy depths of 0.7-1.2 eV were estimated by thermoluminescence measurements (Figure 4c). At increasing temperatures, the defects release energy to the $[SbCl_6]^{3-}$ centers and thus compensate the non-radiative losses. Additionally, while 0D halides undergo significant thermal expansion, XRD analysis revealed that this is mainly accounted for by the elongation of the inter-octahedral distances, while no substantial distortion of the $[SbCl_6]^{3-}$ centers was observed. Thus, the intrinsic rigidity of the emitting sites was considered as the origin of the suppression of non-radiative processes at high temperatures (Figure 4d). Using this anti-TQ metal halide, high-power white light-emitting diodes were fabricated. These maintained stable PL intensity and chromaticity under currents in the 50-2000 mA range (Figure 4e), with performances comparable to those of commercial metal oxide/nitride phosphors[2, 47].

Later on, a work from our groups reported an anti-TQ NIR metal halide phosphor, $Cs_2ZnCl_4$:$xSb^{3+}$, in the form of centimeter-scale crystals[22]. The sample had an PL emission peaked at 745 nm, with a PLQY of up to 75%, and exhibited robust anti-TQ behavior up to 500 K. The anti-TQ behavior of this sample was excitation-dependent, appearing at wavelengths above 370 nm (Figure 4f). Atomistic simulations suggested that this anti-TQ behavior stems from the interplay among host/dopant, dark/bright excited states, phonons, and thermal energy. As shown in Figure 4g, the high-energy photons (e.g. 320 nm, green upward arrow) excite mainly the host, while the low-energy photons (e.g. 390 nm, green upward arrow) directly excite the dopant. Direct guest excitation leads to a higher population of the guest bright state (purple curved line). In this scenario, the limited number of dopant states quickly become saturated, forcing excess carriers into the dark state. At high temperature, the thermally accelerated transition from dark state to bright state compensates the non-radiative loss of the bright state (broad red arrow), a process that should rationalize the robust anti-TQ emission. Instead, when the host is excited, the guest state population is lower, as the host-guest transition is thermally activated. Thus, the carriers recombine before they can overcome the significant energy barrier to the dark state. As a result, the energy transfer from dark state to bright state is not feasible in those excitation conditions. These anti-TQ NIR phosphors were then exploited to fabricate a LED that exhibited stable emission intensity up to 1000 mA, and its high penetration depth was demonstrated on real-life objects.

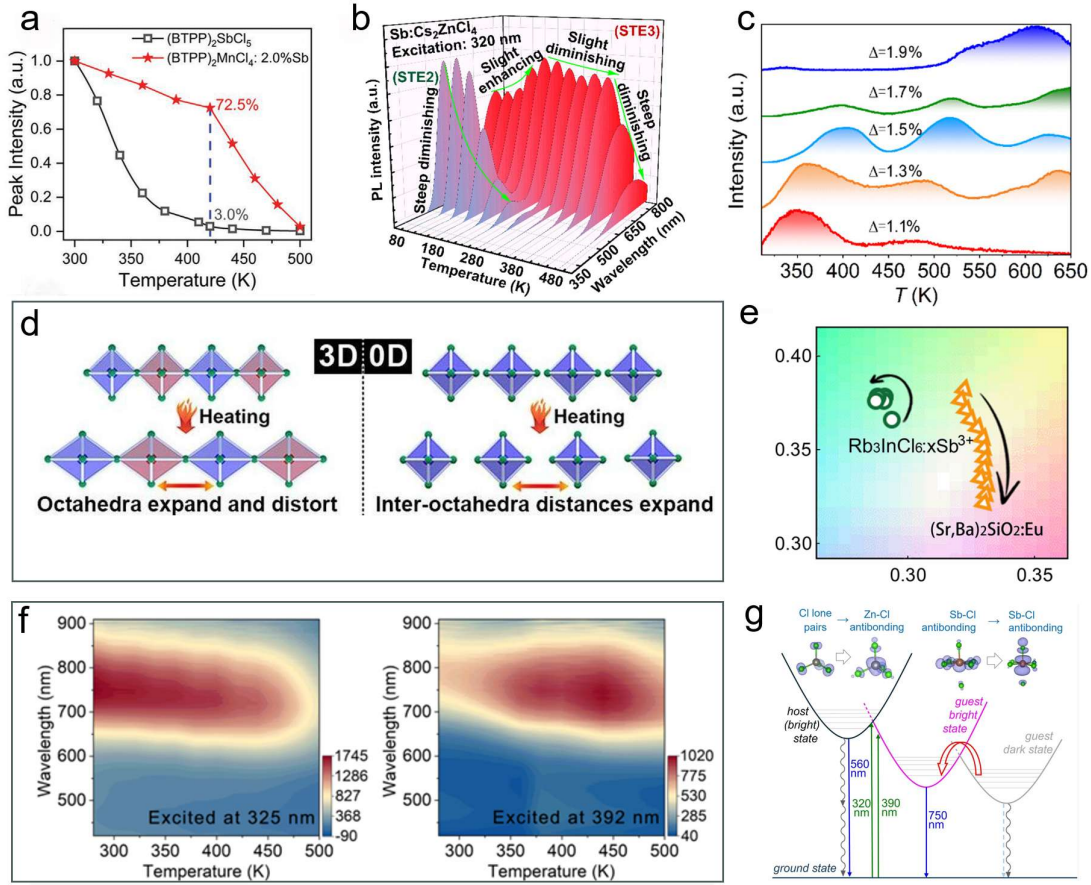

**Figure 4.** $Sb^{3+}$ based anti-TQ metal halides. (a) Temperature dependent PL intensity of $(BTPP)_2SbCl_5$ and $(BTPP)_2SbCl_5$: 2.0%$Sb^{3+}$ at 298-500 K[46]; Reproduced (Adapted) with permission from [46]. Copyright 2022, John Wiley and Sons. (b) PL spectra of the dual emissive $Sb^{3+}$: $Cs_2ZnCl_4$ crystals at 80-480 K[28]. Reproduced (Adapted) with permission from [28]. Copyright 2024, American Chemical Society; (c) Thermoluminescence spectra of $Rb_3InCl_6$:x$Sb^{3+}$ with different deficient values ($\Delta$) of $InCl_3$ precursors during synthesis; (d) Proposed thermal-lattice expansion model for 0D and 3D metal halide frameworks; (e) CIE chromaticity coordinates of the $Rb_3InCl_6$:x$Sb^{3+}$ prototype white light-emitting diodes compared to the commercial green rare-earth oxide phosphor-based white light-emitting diodes, under increasing current (from 50 to 2000 mA)[4]. Reproduced (Adapted) with permission from [4]. Copyright 2024, American Chemical Society; (f) Temperature dependent PL intensities of $Cs_2ZnCl_4$:0.29%$Sb^{3+}$ single crystals at two different excitation wavelengths (left) 325 nm and (right) 392 nm; (g) Proposed model for electronic excitations in $Cs_2ZnCl_4$:$Sb^{3+}$. Green, blue, and curly grey arrows represent, respectively, absorption, radiative recombination, and non-radiative recombination paths. The broad red arrow refers to the thermally assisted energy transfer from the dark to the bright state[22]. Reproduced (Adapted) with permission from [22]. Copyright 2025, John Wiley and Sons.

**2.3. $Mo^{4+}/W^{4+}$ alloy/doping based cases.** In 2022, Liu et al. synthesized $Cs_2MoCl_6$ and $Cs_2WCl_6$ single crystals having NIR emission with a PLQY of 26%[23]. As shown in Figure 5a, the NIR emission intensity of the $Cs_2MoCl_6$ crystals gradually decreased when the temperature increased from -180 to 60 °C, but subsequently increased from 60 to 200 °C, hence demonstrating robust anti-TQ behavior (Figure 5a). Luminescence quenching from -180 to 60 °C was attributed to phonon assisted nonradiative decay, while anti-TQ at high temperature was ascribed to both thermally facilitated STE formation and thermally activated energy transfer from trapped charge carriers to STEs. In 2024, Li et al. synthesized $Mo^{4+}$ doped $Cs_2ZrCl_6$ powders with a broadband NIR emission centered at 970 nm, originating from transitions between the lowest $^1T_2$ state and a manifold of $^3T_1$ states[48]. This NIR phosphor showed mild TQ resistance, maintaining 66.9% of its room temperature PL intensity at 150 °C (Figure 5b), a performance that is comparable to that of most commonly reported $Cr^{3+}$-doped metal oxide phosphors with similar NIR emission[49]. Shubham et al. later prepared alloyed $Cs_2Zr_{1-x}Mo_xCl_6$ powders exhibiting dual emission, from $[ZrCl_6]^{2-}$ STE and from $[MoCl_6]^{2-}$ d-d transitions. Notably, the $[MoCl_6]^{2-}$ NIR emission increased as the temperature increased from 250 to 300 K, thus showing anti-TQ behavior (Figure 5c). The authors proposed that thermal heating softens the lattice, facilitating vibronically coupled d–d transitions and releasing trapped charge carriers[50].

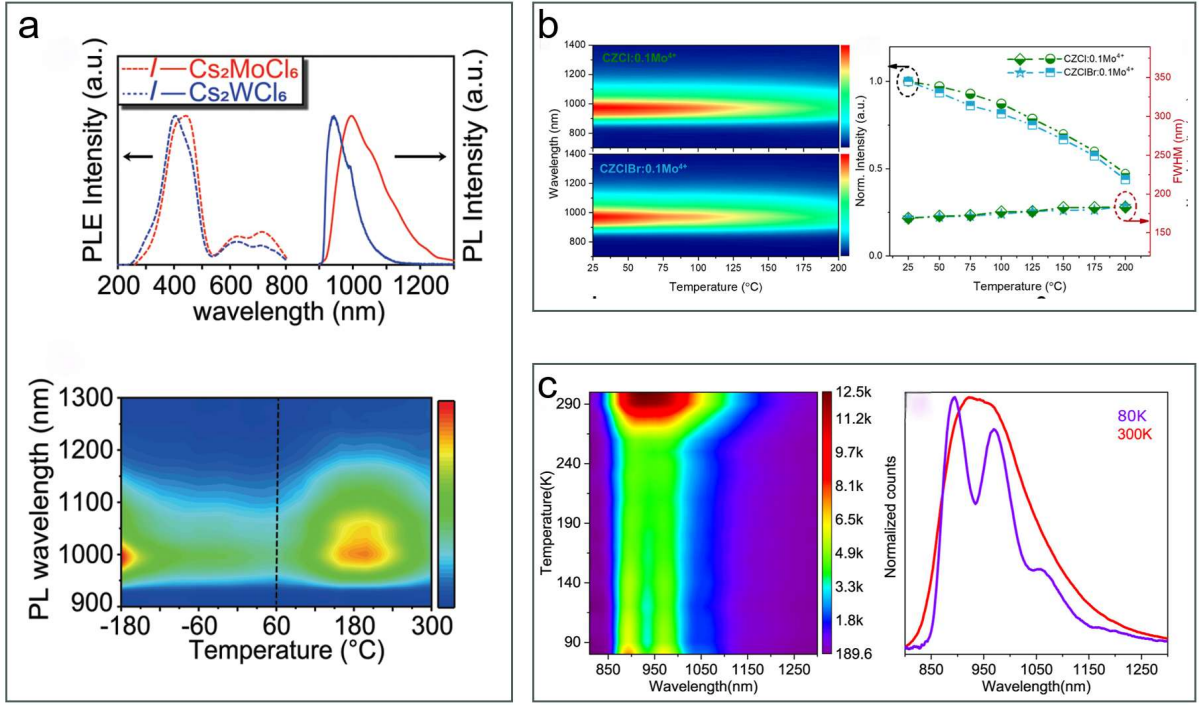

**Figure 5. $Mo^{4+}/W^{4+}$ based anti-TQ metal halides.** (a) PL, PLE and temperature dependent PL spectra of $Cs_2MoCl_6$ single crystals[23]. Reproduced (Adapted) with permission from [23]. Copyright 2022, John Wiley and Sons; (b) Temperature dependent PL spectra of $Cs_2ZrCl_6$: $0.1Mo^{4+}$ and $Cs_2ZrCl_{6-x}Br_x$: $0.1Mo^{4+}$ powders, and their temperature dependent PL intensity and full width at half maximum (FWHM)[48]. Reproduced (Adapted) with permission from [48]. Copyright 2024, American Chemical Society; (c) Temperature dependent alloyed $Cs_2Zr_{1-x}Mo_xCl_6$ powders and its normalized PL spectra at 80 K and 300 K, respectively[50]. Reproduced (Adapted) with permission from [50]. Copyright 2024, American Chemical Society.

**2.4. Hybrid organic-inorganic anti-TQ metal halides.** Some hybrid organic-inorganic metal halides also show anti-TQ behavior through thermal activated energy transfer from metal halides to organic emitters[51]. For example, $(Ph_4P)_2Cd_2Br_6$ maintains stable emission intensity across a wide temperature range from 100 to 320 K ($T_c = \sim 50$ °C)[51]. As this energy transfer process usually involves a singlet-triplet transition, these phosphors also show persistent phosphorescence[20, 51, 52].

### 3. Anti-TQ Metal Halides with Ln ions doping

Yu et al. incorporated $Yb^{3+}$, $Er^{3+}$, and $Ho^{3+}$ into $CsMnCl_3$ single crystals with trigonal structure (space group R-3 m) and achieved TQ resistance above room temperature[53]. In temperature dependent photoluminescence (PL) experiments, the Mn-related emission decreased significantly with increasing temperature, while the $Ln^{3+}$ emission intensity in the near-infrared (NIR) region exhibited zero TQ up to 400 K. As reported in Figure 6a, the $[MnCl_6]^{4+}$ host absorbs UV-visible light and undergoes transitions from $^6A_{1g}(S)$ states to various excited states (blue upward arrow). After that, relaxation can occur through three pathways: (1) radiative transition to the ground state, with emission of red PL ($^4T_{1g}(G) \rightarrow {}^6A_{1g}(S)$, purple downward arrow); (2) crossover to the ground/excited-state crosspoint followed by non-radiative transitions (TQ, dashed arrow); (3) overcoming the energy barrier between the host and dopants, followed by energy transfer to Ln ions (energy transfer, curved red arrow), activating their NIR emission. High temperatures result in thermal quenching by providing sufficient energy to reach the crossover point, thus reducing emission intensity from $[MnCl_6]^{3-}$ states. At the same time, higher temperatures accelerate energy transfer from $[MnCl_6]^{3-}$ to $Ln^{3+}$ dopants, which explains the anti-TQ behavior of these materials.

Zhao et al. and Roh et al. reported anti-TQ $Yb^{3+}$-doped $CsPbX_3$ nanocrystals in 2023[15, 54]. Zhao et al. observed a 2.5-fold enhancement of the Er$^{3+}$ luminescence (peak at 1540 nm) from 298 to 356 K in Yb$^{3+}$/Er$^{3+}$-codoped CsPbCl$_3$ nanocrystals[54], while Roh et al. found a >100 times enhanced Yb$^{3+}$ luminescence intensity from 5K to room temperature in Yb$^{3+}$ doped CsPb(Cl$_x$Br$_{1-x}$)$_3$ nanocrystals. Both groups proposed a thermally assisted quantum cutting process to explain the observed anti-TQ effect[15]. As shown in the left side of Figure 6b, Yb$^{3+}$:CsPbCl$_3$ has E$_g$ > 2 × E$_{f-f}$ and is capable of quantum cutting at all temperatures, with release of excess energy via phonon emission. However, Yb$^{3+}$:CsPbCl$_{3-x}$Br$_x$ (0.5≤ x ≤ 1.00) has E$_g$ ≈ 2 × E$_{f-f}$, so that its lowest donor transition is slightly below the lowest Yb$^{3+}$-Yb$^{3+}$ transition (middle part of Figure 6b). At high temperatures, thermal band-gap widening and line broadening assist these transitions from donor to Yb$^{3+}$-Yb$^{3+}$ and thus improve the quantum cutting process. In contrast, Yb$^{3+}$:CsPbBr$_3$ has E$_g$ < 2 × E$_{f-f}$, so it does not allow a quantum cutting process (right side of Figure 6b). Additionally, heating was also believed to remove excess water molecules from the surface of the nanocrystals and this process was hypothesized to heal the surface traps and further suppress the TQ of Yb$^{3+}$/Er$^{3+}$-codoped CsPbCl$_3$ nanocrystals[54].

Wang et al prepared Sb$^{3+}$/Yb$^{3+}$-codoped Cs$_2$NaHoCl$_6$ single crystals in which the Ho$^{3+}$ luminescence exhibited anti-TQ in the 80-500 K temperature range (Figure 6c)[16]. They attributed anti-TQ to environment-inert 4f-4f transitions. Additionally, they claimed that thermal expansion of the matrix reduced the local concentration of Ho$^{3+}$ ions and diminished the concentration quenching effect caused by non-radiative energy transfer between neighbor dopants. However, this lattice-expansion induced concentration reduction should be minor. For example, based on the temperature dependent XRD, the volume of Rb$_3$InCl$_6$:xSb$^{3+}$ lattice expands only ~3.7% and thus will only result ~3.7% concentration reduction of Sb$^{3+}$ dopants from 300 K to 500 K[4].

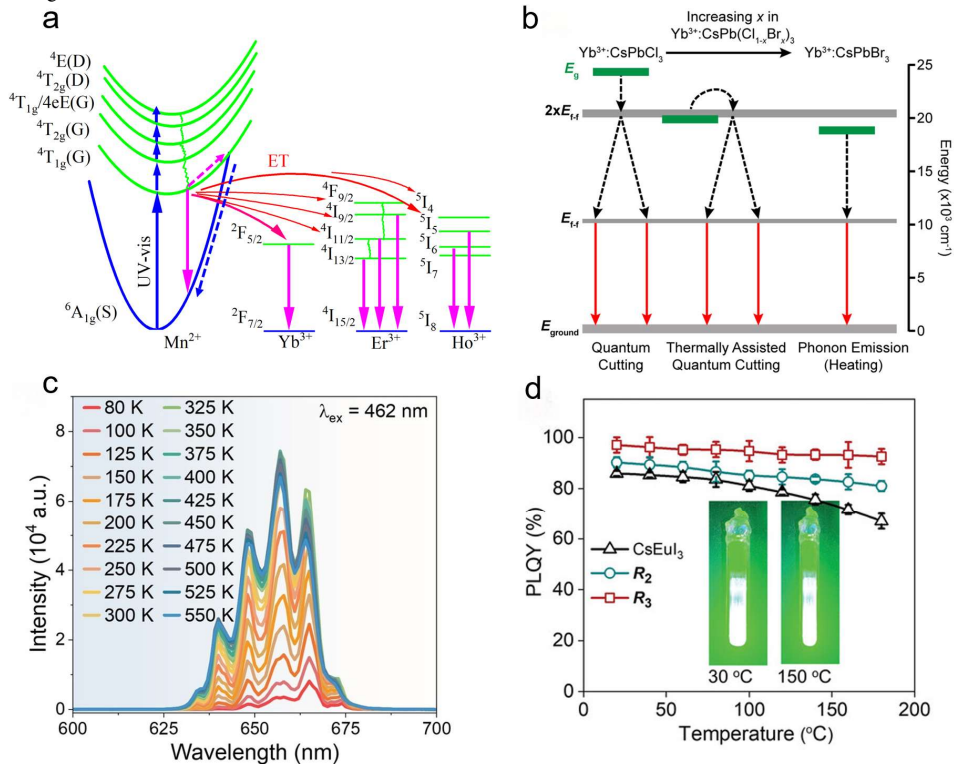

**Figure 6. Ln based anti-TQ metal halides phosphors.** (a) Energy level diagram and electron dynamics of CsMnCl$_3$: Ln$^{3+}$ (Ln = Yb, Er, and Ho) single crystals[53]. Reproduced (Adapted) with permission from [53]. Copyright 2022, Elsevier; (b) Quantum-cutting energy transfer in Yb$^{3+}$:CsPb(Cl$_{1-x}$Br$_x$)$_3$ nanocrystals. The quantum-cutting process is thermally activated[15]. Reproduced (Adapted) with permission from [15]. Copyright 2023, American Chemical Society; (c) Temperature dependent PL spectra of Sb$^{3+}$/Yb$^{3+}$-codoped Cs$_2$NaHoCl$_6$ single crystals in the temperature range from 80 to 550 K[16]. Reproduced (Adapted) with permission from [16]. Copyright 2023, John Wiley and Sons; (d) Temperature-dependent PLQY of 3D (triangle), 1D (circle) and 0D (square) Eu halides powders in the 25–180 °C temperature range[33]. Reproduced (Adapted) with permission from [33]. Copyright 2024, John Wiley and Sons.

Some Ln ions (Eu$^{2+}$, Ce$^{3+}$) undergo 4f-5d transitions whose PL emission is more environment-sensitive compared to the more standard Ln$^{3+}$ case, where only 4f-4f transitions can occur. In 2024, Han et al. developed a series of hybrid organic–inorganic Eu(II)-based halides powders with structural dimensionality ranging from 0D to 1D and 3D[33]. The 0D Eu halide exhibited optimal anti-TQ, with PL decreasing only slightly from 98% to 93% in the 25-180°C temperature range (Figure 6d). These Eu halides feature a narrower emission (FWHM ≈43 nm) compared to commercial metal nitride phosphors (β-SiAlON: Eu$^{2+}$, FWHM ≈54 nm) and their anti-TQ properties are better than those of most lead-based halide perovskites.

## 4. Anti-TQ in Lead Halide Perovskites following Surface Treatments

Lead halide perovskites have been extensively applied in electroluminescence devices, although they suffer from rapid efficiency roll-off at high operating temperatures. To address this, Liu et al. treated $CsPbBr_3$ nanocrystals post-synthesis with dodecyl dimethylammonium fluoride (DDAF, Figure 7a)[17]. The DDAF-treated $CsPbBr_3$ nanocrystals exhibited only slight emission quenching (10% intensity loss) from 298 K to 373 K, compared to nearly complete quenching in pristine nanocrystals (Figure 7a). The authors proposed that the fluorine-rich surface leads to a wide-bandgap type-I 'core-shell' structure, suppressing carrier trapping while improving thermal stability and charge injection. These nanocrystals were incorporated into light-emitting diodes that achieved 19.3% external quantum efficiency (EQE) at 350 cd·m$^{-2}$ and retained nearly 80% of their room-temperature EQE at 343 K. They subsequently tested various anions (OH$^-$, SO$_4^{2-}$, F$^-$) for the $CsPbBr_3$ surface treatment, and discovered that they all lead to the formation of wide-bandgap passivation layers (PbSO$_4$, Pb(OH)$_2$, and PbF$_2$). The results indicated that not only F$^-$ but also SO$_4^{2-}$ ions improve the TQ resistance, preserving 70% of the room temperature PL intensity at 373 K[55]. Wang et al. further incorporated the fluoride treated $CsPbBr_3$ nanocrystals into silica matrices. The resulting fluorine-treated $CsPbBr_3$/silica composites exhibited a high PLQY of 94.5% and preserved 83% of their room temperature PL intensity at 348 K (Figure 7b), plus a complete PL recovery after a thermal cycling (up to 403 K). Also, they retained 95.6% of their initial PL intensity after 1055 h of intense blue light irradiation (350 mW*cm$^{-2}$)[34].

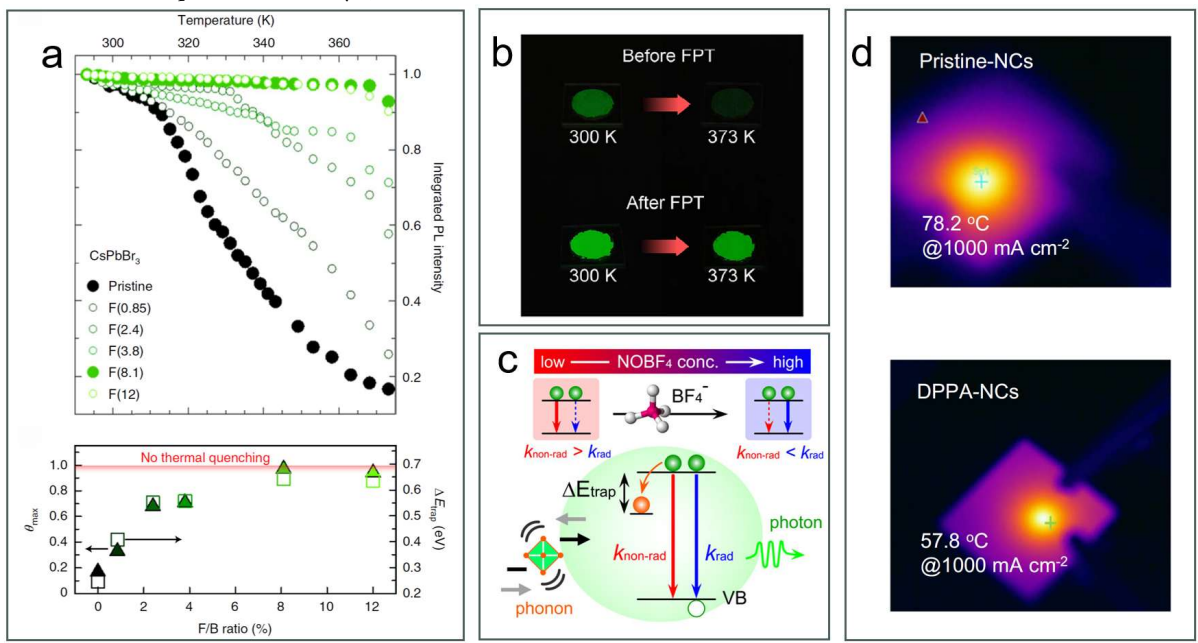

**Figure 7.** Anti-TQ lead halide perovskite nanocrystals with surface treatment. (a) Integrated PL intensity of pristine and fluoride-treated $CsPbBr_3$ nanocrystals at increasing temperatures, from 293 to 373 K, and TQ resistance ($\theta_{max}$) and $\Delta E_{trap}$ of $CsPbBr_3$ nanocrystals as a function of the F/Br ratio[17]. Reproduced (Adapted) with permission from [17]. Copyright 2021, Springer Nature; (b) Comparison of TQ resistance between the untreated $CsPbBr_3$ nanocrystals and fluorine-treated $CsPbBr_3$/silica composites[34]. Reproduced (Adapted) with permission from [34]. Copyright 2025, John Wiley and Sons; (c) The proposed TQ resistance model of NOBF$_4$ treated $CsPbBr_3$ nanocrystals[18]. Reproduced (Adapted) with permission from [18]. Copyright 2024, American Chemical Society; (d) Comparison of the QLED operation temperature between the pristine nanocrystals and the diphenylphosphoryl azide (DPPA) treated ones[56]. Reproduced (Adapted) with permission from [56]. Copyright 2024, Springer Nature.

In 2024, Jeon et al. reported that a NOBF$_4$ treatment improves both PLQY and TQ resistance of CsPbBr$_3$ nanocrystals. In particular, the treated nanocrystals exhibited almost no thermal quenching in the 80–250 K temperature range, while the untreated sample lost 45% of its PL intensity over the same range (Figure 7c)[18]. Dai et al. introduced diphenylphosphoryl azide (DPPA) during the synthesis of CsPb(Br/I)$_3$ nanocrystals[56]. These ligands effectively passivated the surface of the nanocrystals, increasing the PLQY to near unity values and reducing Joule heating. The highly thermally conductive DPPA ligands were hypothesized to transfer heat efficiently and thus reduce the operation temperature of the devices. As shown in Figure 7d, at an applied current of 1000 mA*cm$^{-2}$, the pristine-nanocrystals based QLED featured an operating temperature of 78.2 °C, while the operating temperature of the DPPA nanocrystals based QLED was only 57.8 °C. These thermally optimized QLEDs exhibited an ultra-bright luminance of 390,000 cd*m$^{-2}$, a peak external quantum efficiency of 25%, and an operational half-life of 20 hours.

**In summary, we have provided an overview of a very young field of research, that of anti-TQ metal halide phosphors.** This focus review covers a very short time span, starting from 2017[14] till the present (2025). Although significant progress has been made in this relatively short time span, we can definitely state that research in this area is still in its early stages. An important point that we want to stress in this review is that mainstream anti-TQ metal oxide/nitride phosphors are typically synthesized above 1000°C due to their high lattice formation energies (Table 1). In contrast, softer-lattice metal halides with comparable anti-TQ properties can be synthesized below 200°C or even at room temperature (Table 1), thus offering potential advantages for metal halides in terms of low-temperature processing and energy efficiency.

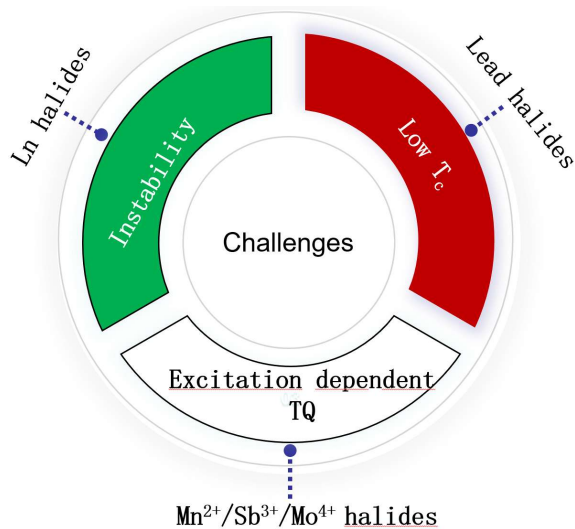

**Figure 8.** The current challenges for three main classes of anti-TQ metal halides discussed in this review.

We have seen how anti-TQ metal halides can be grouped in three classes: (i) phosphors based on metal ions (Mn$^{2+}$, Sb$^{3+}$, Mo$^{4+}$/W$^{4+}$) or on (ii) Ln ions, and (iii) surface-passivated lead halide nanocrystals. For the Ln-based metal halides, the environment-insensitive 4f-4f transitions of Ln$^{3+}$ ions appear promising for anti-TQ phosphors. However, their fixed emission peaks cannot be adjusted by changing the dopant environment. In principle, ions like Eu$^{2+}$ and Ce$^{3+}$, with their hybrid 4f-5d transitions, show greater potential, offering both anti-TQ behavior and tunable emission colors due to the high sensitivity of the 5d orbitals to coordination environments. However, Eu$^{2+}$ halides are very hydrophobic and easy to be oxidized (in this respect they are even more labile than lead halides), and Ce$^{3+}$ halides only show mild TQ resistance. For what concerns metal halides doped with other metal ions, some of them (e.g., Rb$_3$InCl$_6$:Sb$^{3+}$, Cs$_2$ZnCl$_4$:Sb$^{3+}$, Rb$_3$Cd$_2$Cl$_7$:Mn$^{2+}$) exhibit robust zero-thermal quenching up to 500 K and have already been integrated successfully into prototype high-power lighting devices. On a negative side, the anti-TQ mechanisms proposed to date are still unclear and sometimes they contradict each other. For example, some studies suggest that high temperatures distort emission centers and quench STE emission[4, 46], while others propose that they increase available STE states and enhance emission intensity[23, 28, 50]. Also, some of metal halides exhibit excitation dependent anti-TQ behavior[4, 44], which is rarely reported in metal oxides/nitrides anti-TQ phosphors and may hinder their application in WLED. A thorough mechanistic investigation is needed to reconcile these observations. Additionally, low-dimensional metal halides have wide bandgaps and low conductivity, limiting their use in electroluminescent devices.

For what concerns lead halide nanocrystals, they have been successfully employed in electroluminescent devices with superior thermal stability. However, ideal zero-TQ remains unrealized, and most reported nanocrystals show only mild TQ resistance (T$_c$ < 298 K). This is understandable since surface treatments only form a thin layer on the surface of the nanocrystal. Also, this ligand treatment has been reported only for CsPbBr$_3$ nanocrystals, while the possibility to extend anti-TQ behavior also to CsPbI$_3$ and CsPbCl$_3$ nanocrystals is still to be investigated.

Overall, once the above challenges for anti-TQ metal halides are addressed (Figure 8), the application potential on hp-WLED, down-converted displays and high-power lasers will be fully released for this interesting class of materials. In this conclusive section, we also want to lay down some broad guidelines to design an ideal anti-TQ metal halide, based on what we have learned so far: (1) a 0D framework is preferable to a 3D one, since the isolated polyhedra would be less distorted during thermal induced lattice expansion; (2) hetero-valent doping (e.g. Sb$^{3+}$ doped Cs$_2$ZnCl$_4$) could efficiently induce structure defects and thus compensate the non-radiative loss through energy transfer from defects to emitter at high temperature; (3) beyond ligand treatments, researchers should develop additional strategies (e.g., doping or alloying) to further enhance and extend anti-TQ properties of lead halides phosphors.

Finally, we hope that this focus review will stimulate additional studies toward cost-effective, energy-efficient, and multifunctional anti-TQ metal halide phosphors.

AUTHOR INFORMATION

**Corresponding Author**

**Baowei Zhang** – *College of Chemistry, Zhengzhou University, Zhengzhou, Kexue road 450001 China;*

Email: baoweizhang@zzu.edu.cn

**Liberato Manna** – *Nanochemistry, Istituto Italiano di Tecnologia, Genova, Via Morego 30, Italy;*

Email: liberato.manna@iit.it

**Funding**

This work was supported by the National Natural Science Foundation of China (Nos. 22305224), the China Postdoctoral Science Foundation (2022TQ0290), and the European Research Council through the ERC Advanced Grant NEHA (grant agreement n. 101095974).

**Notes**

The authors declare no competing financial interest.

**Biographies**

Baowei Zhang completed his Ph.D. at the Istituto Italiano di Tecnologia. He is currently an associate professor at Zhengzhou University, investigating the synthesis and assembly of semiconductor nanocluster/nanocrystals, and study their optical/mechanical properties.

Liberato Manna completed his Ph.D. at the University of Bari. He is currently head of the Nanochemistry Department and Associate Director at the Istituto Italiano di Tecnologia. His research interests include the synthesis and assembly of nanomaterials, the study of their structural, chemical, and surface transformations, and their applications in energy, photonics, and electronics.

TOC

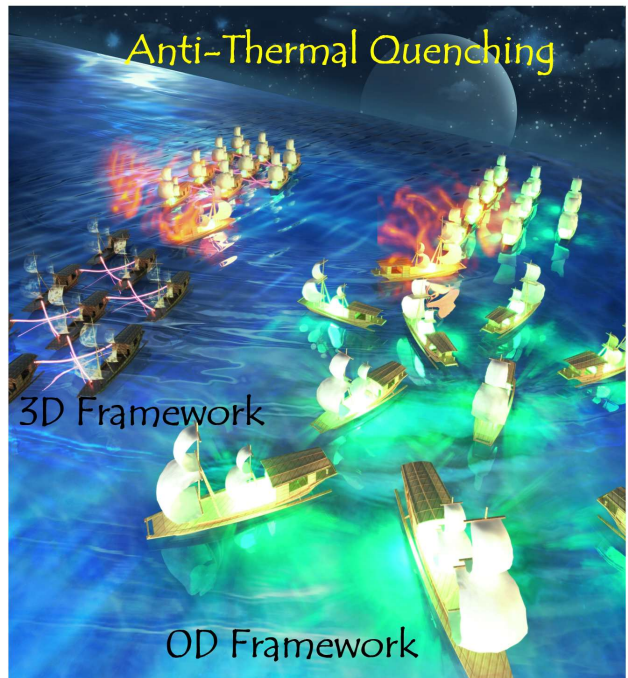

Updated TOC: In a decisive naval battle (named as 'war of Chibi') in three-kingdoms period of China (AD 208–209), the north kingdom had connected each ship by chains (3D framework), to reduce the general seasickness among north people. However, these connected ships were all burned and had no chance to escape after fire attack from the south aliens. If north kingdom would have isolated their ships (0D framework), they would have probably avoided this defeat.